\documentclass[copyright,creativecommons]{eptcs}
\providecommand{\event}{LoCoCo 2010} 
\usepackage{breakurl}             

\title{Handling software upgradeability problems with MILP solvers\thanks{This
work was partially supported by the European Community's 7th Framework Programme (FP7/2007-2013), 
MANCOOSI project, grant agreement n. 214898.}}

\author{Claude Michel \qquad\qquad Michel Rueher
\institute{University of Nice -- Sophia Antipolis / CNRS}
\institute{I3S,
930, Route des Colles - BP 145\\
06903 Sophia Antipolis Cedex\\
\email{\quad Claude.Michel}@i3s.unice.fr\quad\qquad michel.rueher@gmail.com}}

\def\titlerunning{Handling software upgradeability problems with MILP solvers}
\def\authorrunning{C. Michel \& M. Rueher}
\begin{document}
\maketitle

\begin{abstract}
Upgradeability problems are a critical issue in modern operating
systems. The problem consists in finding the ``best'' solution
according to some criteria, to install, remove or upgrade  packages
in a given installation.  
This is a difficult problem: the complexity of the upgradeability
problem is NP complete and modern OS contain a huge number of
packages (often more than 20 000 packages in a Linux
distribution). Moreover,  several optimisation
criteria have to be considered, e.g., stability, memory efficiency, network efficiency. 
In this paper we investigate the capabilities of MILP solvers to handle
this problem. We show that MILP solvers are very efficient when the
resolution is based on a linear combination of the
criteria. Experiments done on real benchmarks show that the best MILP
solvers outperform  CP solvers and that they are significantly
better than Pseudo Boolean solvers. 
\end{abstract}

\section{Introduction}

Upgradeability problems are a critical issue in modern operating
systems.   Indeed, complex software systems are made of numerous interconnected
 components. Free and Open Source
Software (FOSS) distributions  are examples of such systems developed
by distinct individuals  or entities who share their 
work.  FOSS distributions raise difficult problems both for
distribution editors and system administrators. Distributions evolve rapidly by
integrating new versions of software packages that are developed
independently. System upgrades may proceed on different paths
depending on 
the current state of a system and the available software packages.

Installing a software component can be a puzzle because there are
constraints between the different pieces of software (called
packages).   Indeed, open systems
also tend to be much more complex, and therefore some packages may
become incompatible. 

The {\em Mancoosi} project\footnote{See
\url{http://www.mancoosi.org/}}  aims at developing tools for  system
administrators which are faced with choices of upgrade paths, and
possibly with failing upgrades. We
investigated the upgradeability problem in the context of this project.

\subsection{The upgradeability problem}      

The problem consists in finding the ``best'' solution according to
some criteria  to install, remove or upgraded some packages in a given
installation.   This is a difficult problem: the complexity of the
upgradeability problem is at least NP-hard\footnote{The
  installability problem, that's to say ``can we install a package p
  in our system, with a given installation profile P and a package repository 
R?'' is NP-hard \cite{MBD06}, and the installability problem is a
subproblem of the upgradeability problem.} and modern OS contain a
huge number of packages (often more than 20 000 packages in a Linux
distribution).

More formally, the upgradeability problem can be defined in the following way: 
Let $P$ be a set of installed and uninstalled packages,
$p_b$ be a set of packages to be installed, removed or upgraded. The
upgradeability problem consist in finding the best solution $S$
according to some criteria. 

Often, several optimisation criteria have to be considered, e.g.,  stability
(minimise the number of changes in the previous installation),  memory
efficiency (minimise the size of the newly installed packages),
network efficiency (minimise the size of the downloaded packages).   
We consider here that the objective function is defined by a linear combination of such criteria.

\subsection{Contribution}

In this paper we investigate the capabilities of MILP solvers to handle
the Upgradeability problem.   We show that best MILP solvers are very
efficient when the resolution is based on a linear combination of the
criteria.   MILP solvers still behave well with a classical
implementation of a lexicographic order based on mono criterion
solvers.   Experiments done on real problem with at least  20000
packages show that best MILP solvers outperform CP solvers and that
they are significantly better than Pseudo Boolean solvers.        
Very preliminary experiments show that MILP solvers still behave well
with a classical implementation of a lexicographic order, based on mono
criterion solvers.

\subsection{Outline of the paper}    

Section \ref{sec-cudf} introduces
the CUDF, a common upgradeability format \cite{cudf20} which allows to
handle smoothly  variations in the Linux distribution package
description systems. Section \ref{sec-mod} describes the MILP model we
defined to handle the upgradeability problem. Section \ref{sec-exp}
reports results of experiments performed on  $208$ problems ($113$ real problems and $95$ randomly generated problems with a size
ranging from 20000 to  50000 packages). Related works and further
research are discussed in section  \ref{sec-disc}.


\section{The common upgradeability description format} \label{sec-cudf}

An upgradeability problem is fully defined by a set of package descriptions and a set of request descriptions.
However, each Linux distribution uses its own package description system with subtle differences
though most of them derive from the RPM or the debian package formats.
To handle smoothly these variations, a common upgradeability format
(CUDF) \cite{cudf20} defines a  superset
of the various available package descriptions and introduces an uniform
package version numbering.
This section gives the CUDF insights required to understand this paper.
A more complete description of the CUDF can be found in \cite{cudf20}.

A package is defined by its  {\em name} and its  {\em version} (see
figure \ref{cudfex}). An integer denotes the package version with the
convention that they are 
ordered from the lowest up to the highest available version. 
A couple   $< package\;  name , package\; version>$ must be unique in a problem description.

\begin{figure}[!t]
\begin{center}
{\small
\begin{verbatim}
package: car                                                    
version: 1                                         
depends: engine, wheel, door, battery             
installed: true                                    
description: 4-wheeled, motor-powered vehicle      
                                                  

package: gasoline-engine                           package: gasoline-engine  
version: 2                                         version: 1     
provides : engine                                  depends: turbo
conflicts : engine , gasoline-engine               provides: engine
                                                   conflicts: engine, gasoline-engine
                                                   installed: true

package: electric-engine                           package: battery
version: 1                                         version: 3
depends: solar-collector | huge-battery            provides : huge-battery
provides : engine                                  installed : true
conflicts : engine , electric-engine
...

request : 
install : bicycle , electric-engine = 1
upgrade: door
\end{verbatim}
}
\caption{A CUDF example (extracted from \cite{cudf20})\label{cudfex}}
\end{center}
\end{figure}

The depend and conflict fields  describe relationships
of the current package with other packages. The {\em depend} field gives the set of packages required
to install the current package. It is defined by a CNF formula where each package
name can be filtered by an operator on a version to limit the set of
acceptable version for this package, e.g.   electric-engine = 1 or  electric-engine $\ge$ 1.
The {\em conflict} field describes the packages
which conflict with the current package, i.e., the current package cannot be installed 
if any of these packages is installed.
Note that a package might conflict with itself. In such a case, it means that no other version of
the current package can be installed.

The  {\em provide} field describes the set of features provided by the current package.
Here, names  can be virtual package names, i.e.,
names of packages with no available description. 
That way, two different packages with different names can provide the same feature.

The   {\em installed} field is a Boolean which states whether a package is installed or not
in the initial configuration. 

The problem description is defined by  a set of requests  specifying the operations which must
be done on the initial configuration to get the final configuration.
The CUDF format allows three types of operations: install, remove or upgrade.


\section{A MILP model for the upgradeability problem}\label{sec-mod}

The upgradeability problem aims at finding the best solution according to some given criteria.
That is why we investigated the capabilities of MILP to solve this problem.
In other words, we translate the upgradeability problem into
a minimisation problem of a set of binary variables under some integer linear equations
and inequalities. Multicriteria optimisation is  handled through an aggregate function.

\subsection{Constraints}

Modelling the CUDF problem as a linear integer program is quite straightforward.
Each unique couple $<package, version>$ is represented by a binary variable, the value
of which states whether the couple $<package, version>$ is installed or not.
So, the domain of these variables is restricted to $\{0, 1\}$ and solving the problem consists in
finding an assignment of these variables, i.e., determining whether the corresponding couple
$<package, version>$ is installed or not in the final configuration.

A depend field provides a description of the related package dependencies by means of a
conjunction of disjunctions of package names. Assume that $p\_v$, i.e., package $p$ in version $v$,
has the  depend field: 
\begin{displaymath}
 \bigwedge_{i = 1}^{n} p_i\_v_i \wedge \bigwedge_{j = 1}^{m}\bigvee_{k = 1}^{l_m} p_{j,k}\_v_{j,k}
\end{displaymath}
We translate such a formulae into a set of
of integer linear inequalities in two steps:
\begin {enumerate}
\item The first set of conjunctions of the formulae is translated into the following inequality: 
\begin{displaymath}
- n* p\_v + \sum_{i = 1}^{n} p_i\_v_i >= 0
\end{displaymath}
Such an inequality ensures that  all $p_i\_v_i$ are
installed if  $p\_v = 1$, i.e., if $p\_v$ is installed. Of course, the $p_i\_v_i$ can take any
value if $p\_v$ is not installed.
\item   The following integer linear inequality is generated for each disjunction:
\begin{displaymath}
- p\_v + \sum_{k = 1}^{l_m} p_{j,k}\_v_{j,k} >= 0
\end{displaymath}
This inequality ensures that  at least one of the $p_{j,k}\_v_{j,k}$ is installed if $p\_v$ is installed.
\end{enumerate}

\noindent Each conflict field is translated into  the following inequality:
\begin{displaymath}
n' * p\_v + \sum_{p'\_v'\; \in\; {\mathcal Conflict}(p\_v)} p'\_v' <= n'
\end{displaymath}
where ${\mathcal Conflict}(p\_v)$ is the set of packages conflicting with $p\_v$
and $n'$ is the cardinality of ${\mathcal Conflict}(p\_v)$.\\
Such an inequality ensures that  none of the $p\_v$ conflicting packages
can be installed if $p\_v$ is installed.\\

To illustrate this translation process, we provide hereafter part of
the model generated for \verb|gasoline-| \verb|engine_1|, the gasoline-engine
package in version 1 (see  figure \ref{cudfex}).
 \verb|gasoline-engine_1| depends from package \verb|turbo_1| which only exists in version 1.
To ensure that \verb|turbo_1| is installed whenever \verb|gasoline-engine_1| is installed, the following
constraint is generated:
\begin{verbatim}
- gasoline-engine_1 + turbo_1 >= 0
\end{verbatim}
 \verb|gasoline-engine_1| conflicts with any other version of gasoline-engine (like  \verb|gasoline-engine_2|)
as well as with any other package providing the engine feature (like \verb|electric-engine_1| and \verb|electric-| \verb|engine_2|).
To ensure that none of these packages is installed whenever \verb|gasoline-engine_1| is installed, 
the following constraint is generated:
\begin{verbatim}
3 gasoline-engine_1 + gasoline-engine_2 
                    + electric-engine_1 + electric-engine_2 <= 3
\end{verbatim}

The provide field does not directly involve constraint generation. 
As a matter of fact, it is taken into account while managing the
depend or conflict fields through the interpretation of feature names
into set of related package names. For instance, when package {\tt car} asks for
engine in its depend field, the set  \{gasoline-engine version 1, gasoline-engine
version 2, electric-engine version 1\} is substituted to engine.

Once constraints for all the versioned packages have been generated,
the solver handles the problem requests.
Install or remove requests are directly translated by a variable setting
corresponding to the required status in the final configuration.
For instance, constraint $p\_v = 1$ is generated for request {\tt install: p = v}  
and constraint $p\_v = 0$ for request {\tt remove: p = v}.

An upgrade request must ensure that only one version of the upgraded package will be installed
and that the installed package version will be higher or equal to any installed version of
the current package in the initial configuration.
For instance, assume that gasoline-engine has 5 versions ranging from 1 to 5, and that
version 3 is installed in the initial configuration. Then, for  request {\tt upgrade: gasoline-engine},
the solver generates:
\begin{itemize}
\item a constraint that prevents version 1 and 2 to be installed:
\begin{verbatim}
 gasoline-engine_1 + gasoline_engine_2 = 0
\end{verbatim}
\item a constraint to ensure  the uniqueness of the installed version:
\begin{verbatim}
 gasoline-engine_3 + gasoline_engine_4  + gasoline_engine_5 = 1
\end{verbatim}
\end{itemize}

\subsection{Criteria}

We investigated the capabilities of different MILP solvers for the same criterion
which is a variation of the stability criterion, and which is defined  by
the two following criteria: 
\begin{itemize}
\item Criterion (1) : minimize the number of removed functionalities among the installed ones. 
 In other words, we should try to keep
 installed package $p$ if any version of $p$ is installed. This criterion requires the introduction of 
 an additional binary variable $p$ for each package. Remember that the default variables represent the
 status of a couple $<package, version>$, e.g., $p\_v$ which represents the status of
 package $p$ version $v$. 
To make sure that $p$ is true if any version of $p$ is
 installed, and  that $p$ is false otherwise, we add the following constraints:
\begin{displaymath}
- p + \sum_{v_i \in {\mathcal Version}(p)}p\_v_i \geq 0
\end{displaymath}
and
\begin{displaymath}
n'' * p - \sum_{v_i \in {\mathcal Version}(p)}p\_v_i \geq 0
\end{displaymath}
where ${\mathcal Version}(p)$ is the set of versions of $p$,  and $n''$ is ${\;\mathcal Card}({\mathcal Version}(p))$,
the cardinality of ${\mathcal Version}(p)$.
Criterion(1)  is then implemented by:
  \begin{displaymath}
  min \sum_{p\; \in\; F_{\;\mathcal Installed}} - p
  \end{displaymath}
  where $F_{\;\mathcal Installed}$ is the set of installed functionalities.
\item Criterion (2) : minimize the number of modifications,  i.e. if package $p_i$, version $v_i$ is installed 
 keep it installed, if package $p_u$ version $v_u$  is uninstalled keep it
 uninstalled. Criterion (2) is implemented by:
  \begin{displaymath}
  min \sum_{p_i\_v_i\; \in\; P_{\;\mathcal Installed}} - p_i\_v_i + \sum_{p_u\_v_u\; \in\; P_{\;\mathcal Uninstalled}} p_u\_v_u
  \end{displaymath}
  where $P_{\;\mathcal Installed}$ is the set of installed couples $<package,version>$ and $P_{\;\mathcal Uninstalled}$ 
  is the set of uninstalled  couples $<package,version>$.
\end{itemize}
These two criteria are considered in a lexical order so that they can
 be handled independently.
Since the considered solvers optimize only one function -- and to avoid calling them twice for
each problem-- we  aggregated criteria (1) and (2)  in the following way: 
  \begin{displaymath}
  min \sum_{p\; \in\; F_{\;\mathcal Installed}} - {\;\mathcal Card}(P) * p + 
          \sum_{p_i\; \in\; P_{\;\mathcal Installed}} - p_i + \sum_{p_u\; \in\; P_{\;\mathcal Uninstalled}} p_u
  \end{displaymath}
where $P = P_{\;\mathcal Installed} \cup P_{\;\mathcal Uninstalled}$.
Multiplying first criterion coefficients by ${\;\mathcal Card}(P)$ lets any of them have a higher value
than any combination of the second criterion. 
That way, the first criterion  could reach its minimum without being
influenced by the second criterion.  Note, however, that the variables
involved in  the first criterion  are connected to
variables involved in   the second criterion  by  constraints.


\section{Experiments}\label{sec-exp}

Experiments compare $6$ different solvers on a set of $208$ problems
provided by people of Paris Diderot University in the context of the
preparation an international competition of solvers for
package/component installation and upgrade problems (see \url{http://www.mancoosi.org/misc-2010/}).
All the solvers have received
the same set of integer linear constraints. 
Note that the constraints have not been build with a specific solver in mind.

We used  4 MILP solvers and 2  Pseudo Boolean solvers. The tested  MILP solvers are:
\begin{itemize}
\item IBM ILOG CPLEX (version 11.1, see \url{http://www-01.ibm.com/software/integration/optimization/cplex/})  one of the
 best commercial optimization software
 package for solving integer programming problems, linear programming problems, quadratic programming problems, and  convex quadratic constraints;

\item SCIP (version 1.2.0 based on Soplex, see
    \url{http://scip.zib.de/}), one of the
 best non-commercial mixed integer programming solver. 
 branch-cut-and-price. SCIP stands for Solving Constraint Integer
 Programs and combines constraints and LP techniques to solve MILP problems \cite{ABK08};
\item GLPK  (version 4.42, see \url{http://www.gnu.org/software/glpk/}) the GNU Linear Programming
 Kit.  GLPK includes
 a primal and dual simplex method, a primal-dual interior-point method,
a  branch-and-cut method and a stand-alone LP/MIP solver;
\item lp\_solve (version 5.5.0.15, see \url{http://lpsolve.sourceforge.net/}), a Mixed Integer Linear Programming (MILP) solver freely available.
\end{itemize}
The tested  Pseudo Boolean solvers are:
\begin{itemize}
\item WBO \cite{MSP09}, a efficient Pseudo Boolean solver written by Vasco Manquinho,
\item BSOLO (see \url{http://sat.inesc-id.pt/\~vmm/research/index.html}), a Pseudo
 Boolean solver which was first designed to solve instances of the
 Unate and Binate Covering Problems and then adapted to pseubo
 Boolean problem \cite{STM08,RoM09}.
\end{itemize}

As said before, experiments are based on a set of $208$ problems\footnote{Problems are available at \url{http://www.mancoosi.org}} 
ranging from random problems to real problems. All the problems have been solved on an Intel Xeon X5460 quad core @ 3.16Ghz
with 16Gb of memory running under a 64 bit Linux\footnote{All solvers
  were run on a single-threaded mode for the fairness of the comparison.}.
Each table gives the following information
\begin{itemize}
\item ``nb time out'': number of time out (with a time out set to 300s)
\item ``nb failed'': number of problems for which no solution was found
\item ``min time'': minimal time required to solve a problem
\item ``max time'': maximal time required to solve a problem
\item ``geometric mean time'': gives the geometric mean time to solve the problems
\item ``standard deviation'': gives the standard deviation
\item ``total time'': total amount of time required to solve all the problems of the current set
\end{itemize}
Each of the columns reports these information for one of the six
solvers we have compared. 

Here are the details of the experiments we have performed:
\begin{itemize}
\item sets 10orplus (table \ref{tab10orplus}) and 9orless (table \ref{tab9orless}) gather results from two sets 
of real problem provided by Roberto di Cosmo, Mancoosi project leader, who met
some issues while trying to install some packages:
\begin{itemize}
\item the 10orplus set (table \ref{tab10orplus}):  this set of $40$ real problems should involve the installation of more than 10
  packages to fulfill the request. Each problem contains $45998$ packages, $2960$ of them being
  installed and the request consist in the installation of one package.
\item the 9orless set (table \ref{tab9orless}): this set of $38$ real problems should involve the installation of less then $9$
  packages to fulfill the request. The size of these problems is  the
  same than the size of the problems of the 10orplus set.
\end{itemize}
\item caixa set (table \ref{tabcaixa}) gathers results for a set of $45$ real problems coming from Caixa Magica,
  a Linux distributor.
 The size of the problems ranges from $20625$ up to $21045$ packages. $21$ problems have $0$ installed packages,
 $2$ problems, $600$ installed packages and, the last $22$ problems have between $1408$ and $1952$ installed packages.
 Two problems require the installation of $2$ packages while all the other  require the installation of one package.
\item three sets of random problems build from a real installation have also been used. 
  The random part consists in choosing a subset of packages to install or upgrade while each set of random problems
   uses the same initial configuration.
\begin{itemize}
\item rand.biglist set (table \ref{tabrbiglist}) gathers results for a set of $27$ random problems.
    These problems use a huge initial configuration which involves $51449$ packages,
    $551$ of them being installed. One problem requires the upgrade of $551$ packages, another one the upgrade
    of $50$ packages while all the other  ask for the installation of $80$ packages. 
\item rand.newlist set (table \ref{tabrnewlist}) gathers results for a set of $28$ random problems. 
    These problems use also the same  huge initial configuration which involves $51449$ packages
    with $551$ installed packages. However, they only require  to install $30$ packages.
\item rand.smallist set (table \ref{tabrsmallist}) gathers results for a set of $30$ random problems. 
    These problems are based on a smaller initial configuration of $31603$ packages with
    $1145$ installed packages. One problem requires the upgrade of $50$ packages while all the other one ask
    for the installation of $80$ packages.
\end{itemize}
\end{itemize}
To sum up, problem size ranges from $20625$ up to $51449$ packages with none up to $2960$ installed packages
and $1$ up to $551$ packages to install or upgrade.
Table \ref{taballres} provides and overview of the results for all the problems.
\begin{table}[htpb]
\begin{center}
\scriptsize
\begin{tabular}{|l||r|r|r|r|r|r|}
\hline
10orplus (40 problems) & cplex & wbo & scip & glpk & lpsolve & bsolo \\
\hline
nb time out & 0 & 0 & 0 & 39 & 40 & 40 \\
nb failed & 0 & 0 & 0 & 0 & 0 & 0 \\
min time (s) & 5.87 & 36.22 & 25.04 & 282.40 & 300 & 300 \\
max time (s) & 7.83 & 180.14 & 54.50 & 300 & 300 & 300 \\
geometric mean time & 6.25 & 61.45 & 37.26 & 299.55 & 300 & 300 \\
standard deviation & 0.34 & 36.26 & 7.37 & 2.75 & 0 & 0 \\
total time (s) & 250.43 & 2792.24 & 1518.02 & 11982.40 & 12000 & 12000 \\
\hline
\end{tabular}
\caption{Results for 10orplus set of problems}\label{tab10orplus}
\end{center}
\end{table}

\begin{table}[htpb]
\begin{center}
\scriptsize
\begin{tabular}{|l||r|r|r|r|r|r|}
\hline
9orless (38 problems) & cplex & wbo & scip & glpk & lpsolve & bsolo \\
\hline
nb time out & 0 & 0 & 0 & 36 & 38 & 38 \\
nb failed & 0 & 0 & 0 & 0 & 0 & 0 \\
min time (s) & 5.82 & 36.13 & 26.89 & 257.40 & 300 & 300 \\
max time (s) & 7.77 & 58.60 & 52.97 & 300 & 300 & 300 \\
geometric mean time & 6.07 & 41.61 & 36.79 & 298.56 & 300 & 300 \\
standard deviation & 0.29 & 7.14 & 6.57 & 6.93 & 0 & 0 \\
total time (s) & 231.04 & 1602.15 & 1419.80 & 11348.55 & 11400 & 11400 \\
\hline
\end{tabular}
\caption{Results for 9orless set of problems}\label{tab9orless}
\end{center}
\end{table}

\begin{table}[htpb]
\begin{center}
\scriptsize
\begin{tabular}{|l||r|r|r|r|r|r|}
\hline
caixa (45 problems) & cplex & wbo & scip & glpk & lpsolve & bsolo \\
\hline
nb time out & 0 & 0 & 0 & 1 & 11 & 2 \\
nb failed & 32 & 32 & 32 & 32 & 32 & 32 \\
min time (s) & 0.54 & 0.53 & 0.54 & 0.53 & 0.53 & 0.52 \\
max time (s) & 1.33 & 18.93 & 4.89 & 300 & 300 & 300 \\
geometric mean time & 0.85 & 1.69 & 2.2 & 2.6 & 12.07 & 2.96 \\
standard deviation & 0.23 & 5.11 & 1.29 & 44.81 & 129.85 & 64.93 \\
total time (s) & 39.57 & 156.74 & 122.06 & 587.10 & 3944.86 & 1200.87 \\
\hline
\end{tabular}
\caption{Results for caixa set of problems}\label{tabcaixa}
\end{center}
\end{table}

\begin{table}[htpb]
\begin{center}
\scriptsize
\begin{tabular}{|l||r|r|r|r|r|r|}
\hline
rand.biglist (27 problems) & cplex & wbo & scip & glpk & lpsolve & bsolo \\
\hline
nb time out & 0 & 1 & 0 & 14 & 17 & 14 \\
nb failed & 11 & 11 & 11 & 8 & 8 & 11 \\
min time (s) & 1.47 & 1.99 & 4.52 & 2.92 & 21.99 & 2.01 \\
max time (s) & 3.18 & 300 & 193.73 & 300 & 300 & 300 \\
geometric mean time & 2.23 & 69.36 & 9.97 & 58.14 & 153.3 & 30.64 \\
standard deviation & 0.57 & 59.99 & 43.97 & 137.05 & 118.71 & 148.26 \\
total time (s) & 62.39 & 2708.10 & 571.78 & 4631.17 & 5685.08 & 4243.00 \\
\hline
\end{tabular}
\caption{Results for rand.biglist set of problems}\label{tabrbiglist}
\end{center}
\end{table}

\begin{table}[htpb]
\begin{center}
\scriptsize
\begin{tabular}{|l||r|r|r|r|r|r|}
\hline
rand.newlist (28 problems) & cplex & wbo & scip & glpk & lpsolve & bsolo \\
\hline
nb time out & 0 & 0 & 0 & 20 & 25 & 25 \\
nb failed & 3 & 3 & 3 & 3 & 3 & 3 \\
min time (s) & 1.49 & 6.78 & 4.69 & 2.97 & 22.02 & 2.02 \\
max time (s) & 4.08 & 122.02 & 23.10 & 300 & 300 & 300 \\
geometric mean time & 2.83 & 66.68 & 10.89 & 171.36 & 227.14 & 176.49 \\
standard deviation & 0.62 & 27.52 & 3.7 & 95.23 & 85.87 & 92.13 \\
total time (s) & 81.25 & 2105.99 & 322.89 & 7078.26 & 7567.07 & 7506.38 \\
\hline
\end{tabular}
\caption{Results for rand.newlist set of problems}\label{tabrnewlist}
\end{center}
\end{table}

\begin{table}[htpb]
\begin{center}
\scriptsize
\begin{tabular}{|l||r|r|r|r|r|r|}
\hline
rand.smallist (30 problems) & cplex & wbo & scip & glpk & lpsolve & bsolo \\
\hline
nb time out & 0 & 0 & 0 & 2 & 17 & 16 \\
nb failed & 12 & 12 & 12 & 12 & 12 & 12 \\
min time (s) & 0.86 & 1.14 & 2.89 & 1.62 & 8.77 & 1.07 \\
max time (s) & 1.54 & 61.85 & 4.38 & 300 & 300 & 300 \\
geometric mean time & 1.19 & 23.98 & 3.64 & 17.86 & 69.34 & 26.79 \\
standard deviation & 0.22 & 15.95 & 0.44 & 81.98 & 143.42 & 146.98 \\
total time (s) & 36.30 & 912.98 & 110.06 & 1965.43 & 5238.05 & 4890.70 \\
\hline
\end{tabular}
\caption{Results for rand.smallist set of problems}\label{tabrsmallist}
\end{center}
\end{table}

\begin{table}[htpb]
\begin{center}
\scriptsize
\begin{tabular}{|l||r|r|r|r|r|r|}
\hline
All problems (208 problems) & {cplex} & {wbo} & {scip} & {glpk} & {lpsolve} & {bsolo}\\
\hline
{nb time out} &   0  &  1  &  0  &  112  &  148  &  135  \\
{nb fails}    &   58  &  58  &  58  &  55  &  55  &  58  \\
min time (s) & 0.54 & 0.53 & 0.54 & 0.53 & 0.53 & 0.52 \\
max time (s) & 7.83 & 300 & 193.73 & 300 & 300 & 300 \\
geometric mean time & 2.5 & 23.6 & 10.29 & 53.52 & 107 & 53.97 \\
standard deviation & 2.3 & 43.56 & 22.39 & 137.97 & 127.57 & 139.03 \\
{total time (s)} &  700.98 &  10278.20 &  4064.61 &  37653.23  &  45835.06  &  41240.95  \\
\hline
\end{tabular}
\caption{Global results for all the problems}\label{taballres}
\end{center}
\end{table}


\section{Discussion}\label{sec-disc} 

The performances of a state of the art MILP solver such as CPLEX on real
upgradeability problems are really impressive.  This solver is
undoubtedly fast enough to consider its integration in modern
configuration tools. SCIP behaves well too but  performances of the
others MILP solvers are rather disappointing.

The Pseudo Boolean solver WBO behaves well but BSOLO is rather slow. 
However, and contrary to BSOLO, WBO was  slow in proving unsatisfiability.

Experiments with state of art CP solver  like IBM ILOG CP (See
\url{http://www-142.ibm.com/software/products/fr/fr/ilogcp}) where very disappointing: we
could not find any solution for the above-mentioned problems within 300s.

\subsection{Related works}    

The installation problem has been investigated in the EDOS  Project\footnote{See \url{http://www.edos-project.org/}}. This project  aimed at
improving the stability of a distribution from the point of view of
the distribution editor, and not the stability of a particular user
installation.  
SAT based tools have been used to address the installation problem:
e.g., Mancinelli et al formalized the package installation problem as
a SAT problem \cite{MBD06}; 
Josep Argelich and Inês Lynce handled the installability problem as 
a maximum satisfiability (Max-SAT) problem \cite{ArM07}; 
Tucker et al \cite{TSJ07}  addressed the minimal  install/uninstall
problem\footnote{i.e., determine the optimal way to install a new
 package or the minimal number of packages that must be removed from
 a system in order to make a package installable.}. Opium, the
tool they  developed uses Pseudo Boolean and  ILP solvers, and it can
optimize a user-provided objective function, which could for example
state that smaller packages should be preferred to larger ones.

Josep Argelich  and al proposed of Boolean Multilevel Optimization
(BMO) approach to tackle the Upgradeability problem. They used two
different techniques to solve the BMO problem: 1) by iteratively
rescaling the  weights of the MaxSAT formulation;  2) by solving a sequence of
Pseudo Boolean problems. They obtained the best results with WMaxSatz
\cite{ArS09} which could handle problems with up to 4000 packages in a
couple of seconds.  However, this approach could not solve numerous of the
above mentioned problems.

\subsection{Future work} 

Our current work aims at improving the performances of the solvers by
taking advantage of the dependency graph, and by combining CP and MILP
solvers.

Future work concerns also a better handling of preferences, a critical
issue  in  constraint satisfaction and optimization. Note that
 preference-based search algorithms can be generalized  to handle
 multi-criteria optimization \cite{Jun04}.      
Very preliminary experiments show that MILP solvers still behave well
with a classical implementation of a lexicographic order based on mono
criterion solvers. 
However, the time to solve a problem is then proportional to the number of
criteria.


\bibliographystyle{eptcs} 



\end{document}